\documentclass[twocolumn]{aastex631}
\usepackage{mathrsfs}
\usepackage{graphicx}
\usepackage{subfigure}
\usepackage{epstopdf}
\usepackage{amsmath}
\usepackage{amssymb}
\usepackage{hyperref}
\usepackage{color}

\begin{document}
\title{Search for lensing signatures from the latest fast radio burst observations and constraints on the abundance of primordial black holes}

\author{Huan Zhou}
\affiliation{School of Physics and Astronomy, Sun Yat-sen University, Zhuhai, 519082, China}

\author{Zhengxiang Li}
\affiliation{Department of Astronomy, Beijing Normal University, Beijing 100875, China}
\email{zxli918@bnu.edu.cn}

\author{Kai Liao}
\affiliation{School of Physics and Technology, Wuhan University, Wuhan 430072, China}

\author{Chenhui Niu}
\affiliation{National Astronomical Observatories, Chinese Academy of Science, Beijing 100012, China}

\author{He Gao}
\affiliation{Department of Astronomy, Beijing Normal University, Beijing 100875, China}

\author{Zhiqi Huang}
\affiliation{School of Physics and Astronomy, Sun Yat-sen University, Zhuhai, 519082, China}

\author{Lu Huang}
\affiliation{School of Physics and Astronomy, Sun Yat-sen University, Zhuhai, 519082, China}

\author{Bing Zhang}
\affiliation{Nevada Center for Astrophysics, University of Nevada, Las Vegas, NV 89154, USA}
\affiliation{Department of Physics and Astronomy, University of Nevada, Las Vegas, NV 89154, USA}

\begin{abstract}
The possibility that primordial black holes (PBHs) form a part of dark matter has been considered for a long time but poorly constrained over a wide mass range. Fast radio bursts (FRBs) are bright radio transients with millisecond duration. Lensing effect of them has been proposed as one of the cleanest probes for constraining the presence of PBHs in the stellar mass window. In this paper, we first apply the normalised cross-correlation algorithm to search and identify candidates of lensed FRBs in the latest public FRB observations, i.e. $593$ FRBs which mainly consist of the first Canadian Hydrogen Intensity Mapping Experiment FRB catalog, and then derive constraints on the abundance of PBHs from the null search result of lensing signature. For a monochromatic mass distribution, the fraction of dark matter made up of PBHs could be constrained to $\leq87\%$ for $\geq500~M_{\odot}$ at 95\% confidence level by assuming flux ratio thresholds dependent signal-to-noise ratio for each FRB and that apparently one-off events are intrinsic single bursts. This result would be improved by a three times factor when a conventional constant flux ratio threshold is considered. Moreover, we derive constraints on PBHs with a log-normal mass function naturally predicted by some popular inflation models and often investigated with gravitational wave detections. We find that, in this mass distribution scenario, the constraint from currently public FRB observations is relatively weaker than the one from gravitational wave detections. It is foreseen that upcoming complementary multi-messenger observations will yield considerable constraints on the possibilities of PBHs in this intriguing mass window.

\end{abstract}


\section{Introduction}
The cosmological constant plus cold dark matter ($\Lambda$CDM) model has explained the evolution of the universe successfully. The scenario where cold dark matter accounts for about a quarter of the total energy density is well consistent with the large-scale structure observations. However, we still know little about the constituent of dark matter. Primordial black holes (PBHs)~\citep{Hawking1971,Carr1974,Carr1975}, which could form in the early universe from different mechanisms, such as the enhanced curvature perturbations during inflation~\citep{Clesse2015,Pi2018,Ashoorioon2019,Fu2019,Cai2019,Motohashi2020}, bubble collisions~\citep{Hawking1982}, cosmic string~\citep{Hawking1989,Hogan1984}, and domain wall~\citep{Caldwell1996}, have been a source of interest for nearly half a century. One reason for this interest is that only the mass of PBHs can range from the magnitude small enough for Hawking radiation to be important to the level of the black hole in the center of a galaxy. In contrast, astrophysical processes can only form black holes heavier than a particular mass (around three solar mass). 

Moreover, PBHs have also been a source of great astrophysical interest since they are often considered to constitute a part of dark matter. Observational searches for PBHs have been conducted intensively and continuously over several decades. Numerous methods have been proposed to constrain the abundance of PBHs (usually quoted as the fraction of PBHs in dark matter $f_{\rm PBH}=\Omega_{\rm PBH}/\Omega_{\rm DM}$) in various possible mass windows (see \citet{Sasaki2018,Green2020} for a review). These constraints include (in)direct observational effects, such as gravitational lensing~\citep{Niikura2019,Griest2013,Niikura12019a,Tisserand2007,Allsman2001,Zumalacarregui2018,Mediavilla2017,Zhou2021}, dynamical effects on ultrafaint dwarf galaxies~\citep{Brandt2016,Koushiappas2017}, nondetections of stochastic gravitational wave (GW)~\citep{Wang2018,Clesse2017,Chen2020,Luca2020, Gert2020}, disruption of white dwarfs~\citep{Graham2015}, null detection of scalar-induced GW~\citep{Chen2019b}, and the effect of accretion via cosmic microwave background observations~\citep{Haimoud2017,Aloni2017,Chen2016,Poulin2017,Bernal2017}. Constraints on the PBHs lighter than $\sim10^{15}~\rm g$ that have already evaporated by the Hawking radiation can be indirectly derived from certain features in the extragalactic and Galactic $\gamma$-ray backgrounds~\citep{Carr2016,DeRocco2019,Laha2019,Laha2020a,Dasgupta2020}. In addtion to these available probes, some other constraints from the near future observations have been proposed, such as gravitational lensing of GW~\citep{Jung2019,Liao2020a,Diego2020,Urrutia2021}, gamma-ray bursts~\citep{Ji2018}, and 21 cm signals~\citep{Hektor2018,Clark2018,Halder2020}. The mass range $1-100~M_{\odot}$ is now of special interest in view of the recent detection of GW from binary black hole merger~\citep{Abbott2016}. Detection of GW bursts from merges of compact object binaries is one of the most promising ways to study the mass distribution of PBHs. Meanwhile, the abundance of PBHs in the mass range $1-100~M_{\odot}$ also can be well constrained from gravitational lensing effect of prolific transients with millisecond duration, i.e. fast radio bursts (FRBs). 

FRBs are brief (few millisecond) but very bright radio wave pulses and almost all of them have been observed at extragalactic distances \citep{Lorimer2007,Thornton2013,Petroff2015,Petroff2016,Cordes2019,Zhang2020}. According to the detection rate and the field of view of radio telescopes, a high event rate of this kind of mysterious phenomenon ($\sim10^3$ to $10^5$ sky$^{-1}$day$^{-1}$) has been inferred \citep{Thornton2013,Champion2016,2021ApJ...909L...8N}. Although the radiation mechanism and progenitors of these mysteries are still intensively debated \footnote{The recent detection of a Galactic FRB in association with a soft gamma-ray repeater suggests that magnetar engines can produce at least some (or probably all) FRBs \citep{Zhang2020,CHIME/FRB2020,Bochenek2020,Lin2020}.}, observed dispersion measure (DM) of them, which is proportional to the number density of free electron along the line of sight of radio emission, have been explored as promising probes for constraining cosmological models \citep{Gao2014,Zhou2014,Walters2018,Zhao2020}, baryon census \citep{Deng2014,Macquart2020,Li2019,Li2020}, and reionization history of universe \citep{Linder2020,Bhattacharya2020,Beniamini2021}. Moreover, due to prominent observational features including short duration and precision localization for both repeating and apparent one-off FRBs~\citep{Chatterjee2017,Prochaska2019b}, gravitational lensed FRBs have been widely proposed for cosmological and astrophysical studies, such as millilensed lensed FRBs for probing compact dark matter \citep{Munoz2016,Wang2018a,Liao2020,Laha2020,Katz2020}, galaxy lensing time delay variations for probing the the motion of the FRB source \citep{Dai2017}, and time delay distances of strongly lensed FRBs for precisely measuring the expansion rate and curvature of the universe \citep{Li2018,Wucknitz2020}. Recently, one of the most exciting things is the publication of the first Canadian Hydrogen Intensity Mapping Experiments (CHIME) FRB catalog~\footnote{https://www.chime-frb.ca/catalog}~\citep{CHIME/FRB2021}, which makes up of a considerable portion of currently available FRBs.

In this paper, we propose the method of cross-correlation function to search and identify the candidate of lensed FRB and apply the method proposed by \citet{Munoz2016} to constrain the abundance of PBHs with the latest $593$ FRB observations. In addition to the monochromatic mass distribution used in previous studies~\citep{Munoz2016,Liao2020,Laha2020}, we also investigate the results with the popular log-normal mass distribution, which can be compared with constraints from the latest GW catalog.

This paper is organized as follows: we introduce the redshift and duration information of the latest FRBs, review the theory of FRBs lensing, and present the method of identifying lensing signatures in Section 2. In section 3, we apply the method to the latest FRB observations and yield results. In addition, we will compare the results of FRBs with current constraints from GW data; Conclusions and discussions are presented in Section 4.

\section{Methods}
In this section, we briefly introduce the current status of FRB observations, review the FRB lensing theory, and present the method of searching and identifying lensing signatures.

\subsection{Fast Radio Burst Observations}
The number of verified FRBs is increasing rapidly at the moment owing to services of several wide-field radio telescopes, such as the CHIME, the Australian Square Kilometre Array Pathfinder (ASKAP), and the Deep Synoptic Array (DSA). In particular, the CHIME/FRB Collaboration has recently released a catalog of 535 FRBs detected in less than one year (2018 July 25 to 2019 July 1)~\citep{CHIME/FRB2021}. In this catalog, there are 61 bursts from 18 previously reported repeaters. The first large sample observed in a single survey with uniform selection effects is of great value for facilitating comparative and absolute studies of the FRB population. All these FRBs together with bursts detected by other facilities have been collected and compiled by the Transient Name Sever (TNS)~\footnote{https://www.wis-tns.org}. At this moment, there are about 593 independent events publicly available.

\begin{figure}[ht!]
    \centering
     \includegraphics[width=0.4\textwidth, height=0.3\textwidth]{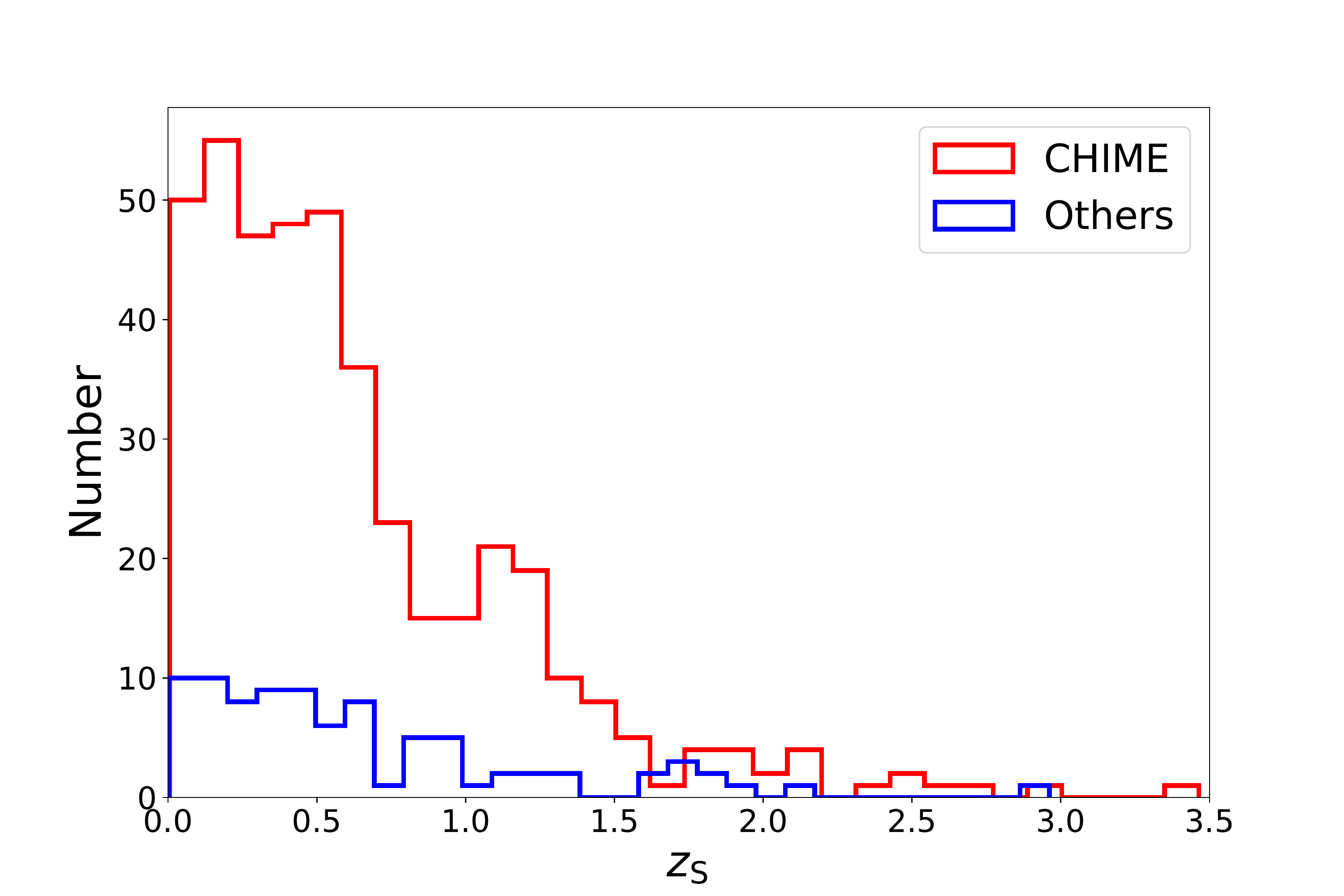}
     \caption{Distribution of inferred redshifts of 593 FRBs, including 492 FRBs from CHIME and 101 FRBs from other facilities.}\label{fig1}
\end{figure}

\begin{figure}[ht!]
    \centering
     \includegraphics[width=0.4\textwidth, height=0.3\textwidth]{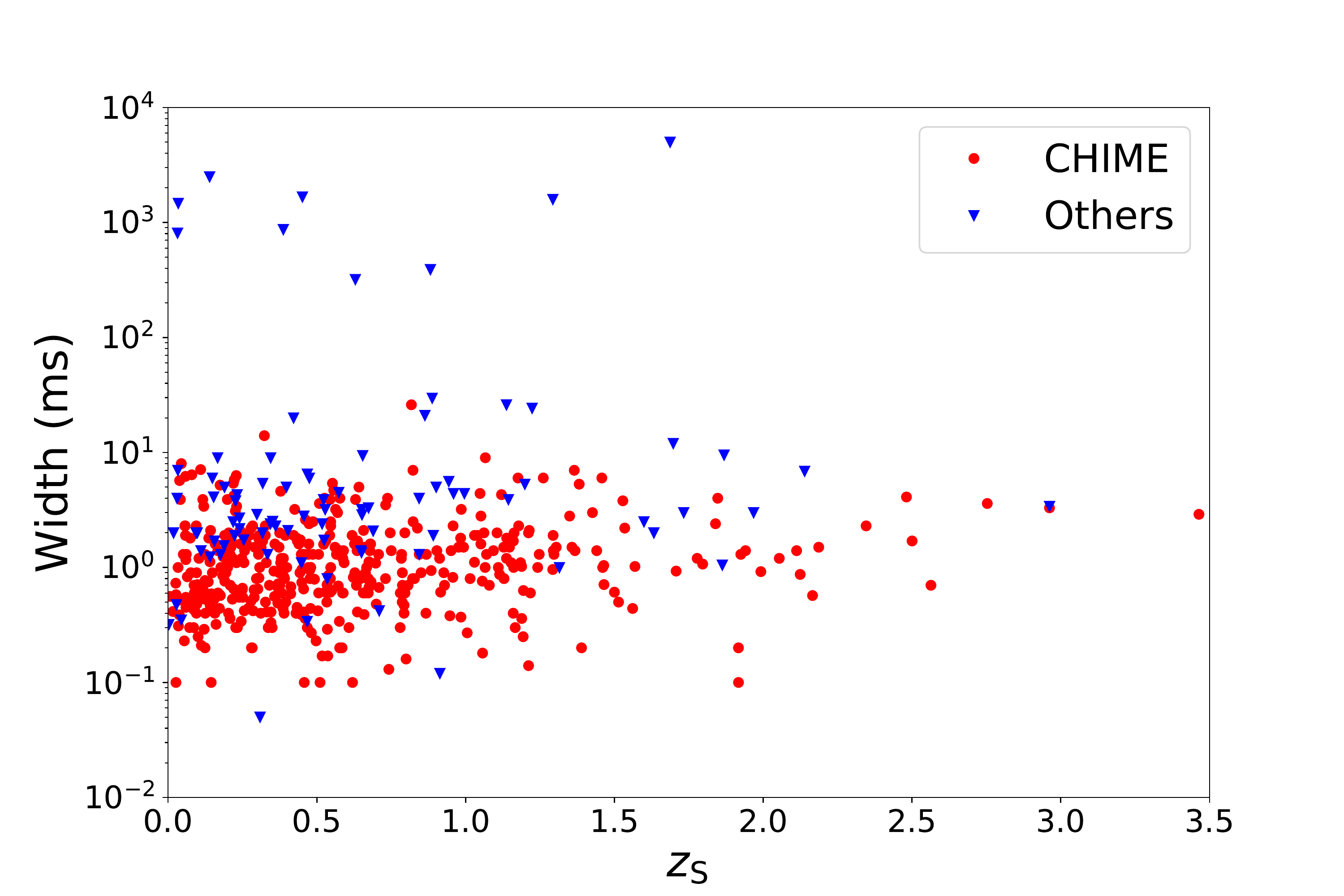}
     \caption{Two-dimensional distribution of widths and inferred redshifts.}\label{fig2}
\end{figure}

For a detected FRB, one of the most important observational features is the DM, which is theoretically defined as the integration of the election number density along the traveled path of the radio pulse and, in observations, obtained by measuring the delayed arrival time of two photons with different frequencies. From observed DMs of the first several bursts which were poorly localized then~\citep{Lorimer2007,Thornton2013}, cosmological origins of this kind of mysterious flashes was inferred. And this inference has been subsequently confirmed by the localization of the first repeater FRB20121102A to a nearby dwarf galaxy~\citep{Tendulkar2017,Chatterjee2017,Marcote17}. Therefore, the distance and the redshift can be roughly derived from the observed DM of a detected FRB which is usually decomposed into:
\begin{equation}\label{eq1}
{\rm DM}={\rm DM_{MW}}+{\rm DM_{IGM}}+\frac{\rm DM_{\rm host}+DM_{\rm src}}{1+z},
\end{equation}
where ${\rm D_{MW}}$ is the contribution from the Milky Way, ${\rm DM_{host}}$ and ${\rm DM_{src}}$ are contributions from FRB host galaxy and source environment, respectively. Here, we conservatively adopt the maximum value of ${\rm DM_{host}}+{\rm DM_{src}}$ to be 200 $\rm pc/cm^{3}$, which corresponds to the minimum inference of redshift for all host galaxies. In addition, the ${\rm DM_{IGM}}-z$ relation is given by~\citet{Deng2014}, approximately ${\rm DM_{IGM}}\sim855z ~\rm pc/cm^3$~\citep{Zhang2018}, with the consideration of He ionization history and the fraction of baryon in intergalactic medium (IGM) $f_{\rm IGM}$ being $0.83$. This relation is statistically favored by the five localized FRBs available at that time~\citep{Li2020}. Inferred redshifts for currently public FRBs from different facilities are shown in Figure~\ref{fig1}. Another important observational feature of FRBs for lensing scales in our following analysis is the short duration. Therefore, we collect the pulse widths of all available FRBs and present them in Figure~\ref{fig2}.

\subsection{Lensing of Fast Radio Bursts}
In~\citet{Munoz2016}, it was first pointed out that lensing effect of FRB can be used to probe PBHs with mass as small as $20~M_{\odot}$. We can take the PBH as a point mass whose Einstein radius is given by:
\begin{equation}\label{eq2}
\theta_{\rm E}=2\sqrt{\frac{GM_{\rm PBH}}{c^2D}}\approx (3\times10^{-6})^{''}\bigg(\frac{M_{\rm PBH}}{M_{\odot}}\bigg)^{1/2}\bigg(\frac{D}{\rm Gpc}\bigg)^{-1/2},
\end{equation}
where $G$ and $c$ denote the gravitational constant and the speed of light, respectively. In addition, $D=D_{\rm L}D_{\rm S}/D_{\rm LS}$ is effective lensing distance, where $D_{\rm S}$, $D_{\rm L}$, and $D_{\rm LS}$ represent the angular diameter distance to the source, to the lens, and between the source and the lens, respectively. Although the spatial resolution in radio observation could reach a high level $\sim(10^{-2})^{''}$, it is still insufficient to distinguish split image for $M_{\rm PBH}<10^8~M_{\odot}$. However, one can directly measure the time delay between lensed signals. The formula of time delay is determined by:
\begin{equation}\label{eq3}
\begin{split}
\Delta t(M_{\rm PBH},z_{\rm L},y)=\frac{4GM_{\rm PBH}}{c^3}\big(1+z_{\rm L}\big)\\
\bigg[\frac{y}{2}\sqrt{y^2+4}+\ln\bigg(\frac{\sqrt{y^2+4}+y}{\sqrt{y^2+4}-y}\bigg)\bigg],
\end{split}
\end{equation}
where the dimensionless impact parameter $y=\beta/\theta_{\rm E}$ stands for the relative source positions, $z_{\rm L}$ is the lens redshift. $\Delta t$ must be larger than the width ($w$) of the observed signal. This requires $y$ larger than a certain value $y_{\rm min}(M_{\rm PBH},z_{\rm L},w)$ according to Equation~\ref{eq3}. The lensing cross section due to a PBH lens is given by:
\begin{equation}\label{eq4}
\begin{split}
\sigma(M_{\rm PBH}, z_{\rm L}, z_{\rm S}, w)=\frac{4\pi GM_{\rm PBH}D_{\rm L}D_{\rm LS}}{c^2D_{\rm S}}\\
[y^2_{\rm max}(R_{\rm f,max})-y^2_{\rm min}(M_{\rm PBH},z_{\rm L},w)].
\end{split}
\end{equation}
The maximum value of normalized impact parameter can be found by requiring that the two lensed images is greater than some reference value of flux ratio $R_{\rm f}$:
\begin{equation}\label{eq5}
y_{\rm max}(R_{\rm f,max})=\sqrt{\frac{1+R_{\rm f,max}}{R_{\rm f,max}^{0.5}}-2}.
\end{equation}
The reference value is usually set as a constant $R_{\rm f,max}=5$~\citep{Munoz2016,Liao2020,Laha2020}. Recently, \citet{Krochek2021} suggested that this quantity should be dependent on the signal-to-noise ratio (SNR) of each observed FRB. This is reasonable since both SNR and flux ratio are crucial for identifying a lensed FRB event. Therefore, these two options are considered in our following analysis. For a given source, the lensing optical depth due to a single PBH is:
\begin{equation}\label{eq6}
\begin{split}
\tau(M_{\rm PBH},f_{\rm PBH},z_{\rm S},w)=\int_0^{z_{\rm S}}d\chi(z_{\rm L})n_{\rm L}(f_{\rm PBH})(1+z_{\rm L})^2\\
\sigma(M_{\rm PBH},z_{\rm L},z_{\rm S}, w)=\frac{3}{2}f_{\rm PBH}\Omega_{\rm DM}\int_0^{z_{\rm S}}dz_{\rm L}\frac{H_0^2}{cH(z_{\rm L})}\\\frac{D_{\rm L}D_{\rm LS}}{D_{\rm S}}
(1+z_{\rm L})^2[y^2_{\rm max}(R_{\rm f,max})-y^2_{\rm min}(M_{\rm PBH},z_{\rm L},w)],
\end{split}
\end{equation}
where $n_{\rm L}$ is the comoving number density of the lens, $H_0$ is the Hubble constant at present universe, $H(z_{\rm L})$ is the Hubble function at $z_{\rm L}$, $f_{\rm PBH}$ represents the fraction of PBHs in dark matter, and $\Omega_{\rm DM}$ is the present density parameter of dark matter. Now, for a given distribution function $N(z_{\rm S})$ of FRBs, we can calculate their integrated lensing optical depth $\bar{\tau}(M_{\rm PBH},f_{\rm PBH},w)$ as:
\begin{equation}\label{eq7}
\bar{\tau}(M_{\rm PBH},f_{\rm PBH},w)=\int dz_{\rm S}\tau(M_{\rm PBH},f_{\rm PBH},z_{\rm S},w)N(z_{\rm S}).
\end{equation}
If we observe a larges number of FRBs, $N_{\rm FRB}$, then the number of FRBs that will be lensed is:
\begin{equation}\label{eq8}
N_{\rm lensed~FRB}=(1-e^{-\bar{\tau}(M_{\rm PBH},f_{\rm PBH},w)})N_{\rm FRB}.
\end{equation}
If none of the FRBs is found to be lensed, then the fraction of dark matter in the form of PBHs can be estimated in Equation~\ref{eq8}. Now, the newest event number is $593$, which holds a statistical meaning. According to the definition, the expected number of lensed FRBs can be approximated to the sum of the lensing optical depths of all FRBs ($\tau_i\ll1$):
\begin{equation}\label{eq9}
N_{\rm lensed~FRB}=\sum_{i=1}^{N_{\rm total}}\tau_i(M_{\rm PBH},f_{\rm PBH},z_{S,i},w_i).
\end{equation}
The above formalism is valid if the mass distribution of the PBHs follows monochromatic mass function. It has been theoretically shown that PBHs can also follow an extended mass distribution function~\citep{Bellomo2018,Carr2017,Laha2020}. We will consider the log-normal mass function of PBHs, and log-normal mass function is parametrized as:
\begin{equation}\label{eq10}
P(m,\sigma,m_{\rm c})=\frac{1}{\sqrt{2\pi}\sigma m}\exp\bigg(-\frac{\ln^2(m/m_{\rm c})}{2\sigma^2}\bigg),
\end{equation}
where $m_{\rm c}$ and $\sigma$ give the peak mass of $mP(m)$ and the width of mass spectrum. This mass function is often a good approximation if PBHs result from a smooth symmetric peak in the inflationary power spectrum. Therefore, it can be representative of a large class of extend mass functions. For a log-normal mass distribution, the lensing optical depth can be expressed as:
\begin{equation}\label{eq11}
\begin{split}
\tau(f_{\rm PBH},z_{\rm S}, w, \sigma,m_{\rm c})=\int dm\int_0^{z_{\rm S}}d\chi(z_{\rm L})n_{\rm L}(f_{\rm PBH})\\
(1+z_{\rm L})^2\sigma(m,z_{\rm L},z_{\rm S}, w)P(m,\sigma,m_{\rm c})=\frac{3}{2}f_{\rm PBH}\Omega_{\rm DM}\\
\int dm\int^{z_{\rm S}}_0dz_{\rm L}\frac{1}{\sqrt{2\pi}\sigma m}\exp\bigg(-\frac{\ln^2(m/m_{\rm c})}{2\sigma^2}\bigg)\frac{H_0^2}{cH(z_{\rm L})}\\
\frac{D_{\rm L}D_{\rm LS}}{D_{\rm S}}(1+z_{\rm L})^2[y^2_{\rm max}(R_{\rm f,max})-y^2_{\rm min}(m,z_{\rm L},w)].
\end{split}
\end{equation}
This calculated value of the optical depth can be used to get the integrated optical depth in Equation~\ref{eq7}.

\subsection{Searching and identifying lensing signatures}
An FRB strongly lensed by a lens mass greater than $\sim20~M_{\odot}$ would be separated into two images with observable time delay $\gtrsim$ few milliseconds and the light curve of it will appear as two distinct peaks. In this case, the lensing time delay would be comparable with or greater than the duration of the burst and can be read manifestly. To detect this, we define the normalized cross-correlation (NCC) of two peaks~\citep{Bracewell1986,Liyang1993,Liwang1996,Gonzalez2002,Hirose2006}:
\begin{equation}\label{eq12}
C(\delta t)=\frac{\int dtI_1(t)I_2(t-\delta t)}{\sqrt{\int dt I^2_1(t)}\sqrt{\int dtI^2_2(t-\delta t)}},
\end{equation}
where $I_1(t)$ and $I_{2}(t)$ are intensities of the two light curve peaks. They consist of contributions of both signal $S_i$ and noise $N_i$, i.e. $I_i=S_i+N_i$. In a specific lensing configuration (high signal-to-noise ratio, e.g. $\rm SNR\gtrsim10$ which is necessary to confirm an FRB signal), the light curve of the first peak $I_1(t)$ should be proportional to the second peak $I_2(t)$ with time delay $\Delta t$ and flux ratio $R_{\rm f}$:
\begin{equation}\label{eq13}
I_1(t)\propto R_{\rm f}I_2(t-\Delta t).
\end{equation}
It is obvious that, for a lensed FRB, NCC will exhibit spikes at different frequency bins with $C(\delta t=\Delta t)\backsimeq1$. Although for peaks with high SNR, noise $N_i$ always make $C(\delta t)$ smaller than unity, we would search the lensed candidates of $C(\delta t=\Delta t)\backsimeq1$ in confirmed FRB signals with considerable SNR.

\begin{figure*}
    \centering
     \includegraphics[width=0.3\textwidth, height=0.4\textwidth]{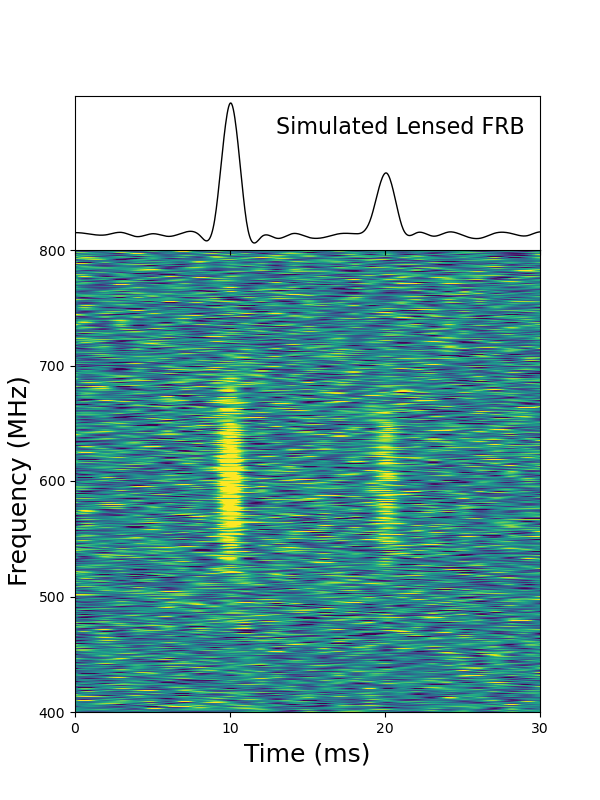}
     \includegraphics[width=0.4\textwidth, height=0.4\textwidth]{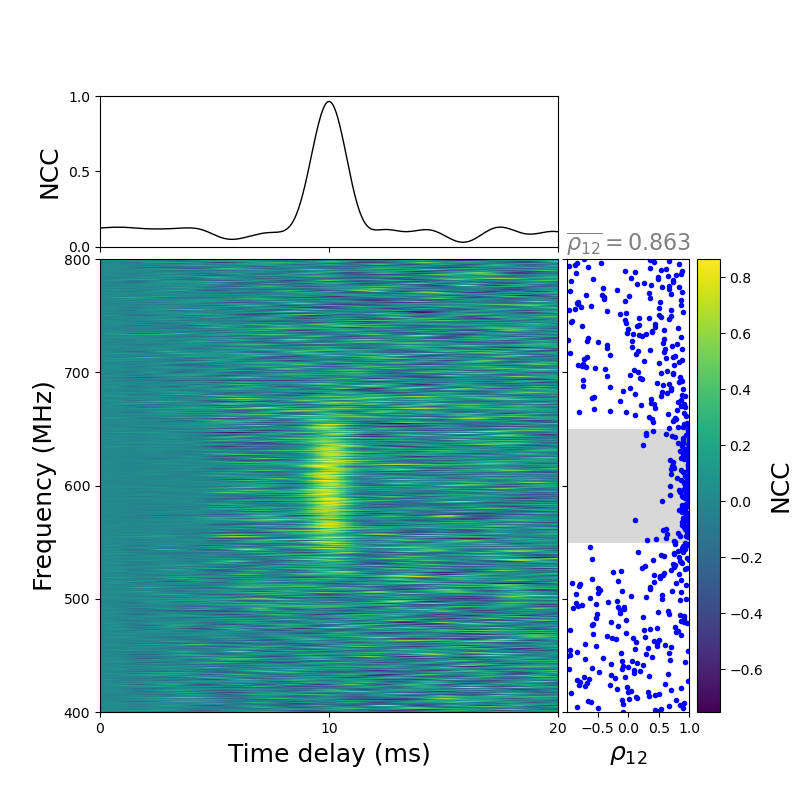}
     \caption{\textbf{Left panel:} the dynamic spectrum of a simulated lensed FRB signal. \textbf{Right panel:} The subfigure in the middle represents the two-dimensional NCC analysis at different frequency bins. The subfigure in the upper left represents the NCC analysis of the light curve of the two peaks in the left panel. The subfigure on the right is the PCC of two pulses in different frequency bins, and $\overline{\rho_{12}}$ represents the average of PCC in the gray shade frequency range. }\label{fig3}
\end{figure*}

To test the validity of this method, we first simulate a positive case (i.e. a typical lensed signal) for the above-mentioned NCC analysis. In the simulation, we set the frequency range of $400$-$800~\rm MHz$ with 16384 frequency channels and the time resolution being $0.98~\rm ms$, which are consistent with characteristics of recently released CHIME FRBs. The simulated de-dispersed bursts that follow the Gaussian profile and its background is injected with Gaussian white noise. We generate two lensed bursts that have the same frequency emission band and pulse structure but different arriving time and intensities. In addition, the SNR of mocked signals is roughly consistent with the averaged level of the CHIME FRB observations. The frequency-time (``waterfall") plot of the mocked signal is shown in the left panel of Figure~\ref{fig3}. Next, we carry out NCC analysis for the mocked pulses and the result is presented in the right panel. As expected, for a typical lensed FRB signal, the values of NCC peak the same time delay in different frequency bins, which is consistent with the prediction of gravitational lensing theory. Moreover, the lensing time delay of two pulses can be obtained as $\Delta t=10\rm~ms$ with flux ratio $R_{\rm f}=2$ and we can infer the redshifted lens mass $M_z=M_{\rm PBH}(1+z_{\rm L})=725.2~M_{\odot}$ based on equation~\ref{eq3}. Finally, we quote the Pearson correlation coefficient (PCC:  $\rho_{12}$)~\citep{Pearson1896,Dunn1974,Rodgers1988} to evaluate the degree of correlation between two peaks (peak1,~peak2). In relation to the range of values of $\rho_{12}$, we can distinguish the following cases:
\begin{itemize}
\item $\rho_{12}\sim(0.7,~1)$, it testifies a strong positive correlation of the dependent peak1 with the independent peak2.
\item $\rho_{12}\sim(0.3,~0.7)$, it testifies a moderate positive correlation of the dependent peak1 with the independent peak2.
\item $\rho_{12}\sim(0,~0.3)$, it testifies a weak positive correlation of the dependent peak1 with the independent peak2.
\end{itemize}
The value of $\rho_{12}$ also could be negative and the magnitude of it testifies corresponding degree of negative correlation. The PCCs for each frequency bin of the mocked signal are plotted in the right panel of Figure \ref{fig3}. Since the PCC value is sensitive to the level of SNR, a mean of PCCs ($\overline{\rho_{12}}$) for frequency bins with SNR larger than a reasonable threshold is derived and also shown in the right panel of Figure \ref{fig3}. For the simulated lensed signal, we obtain $\overline{\rho_{12}}=0.863$ and it testifies a strong positive correlation between these two peaks. That is, lensing signatures could be successfully identified by this algorithm. 

\begin{figure*}
    \centering
     \includegraphics[width=0.3\textwidth, height=0.4\textwidth]{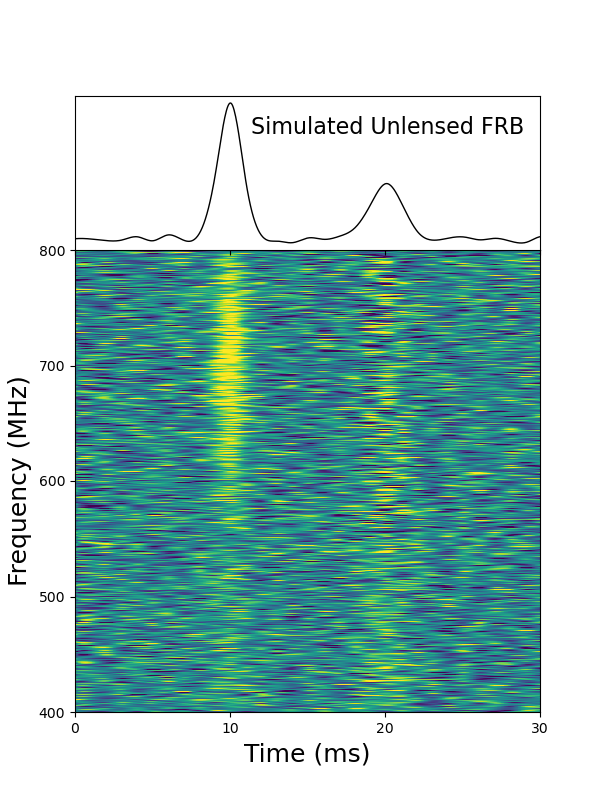}
     \includegraphics[width=0.4\textwidth, height=0.4\textwidth]{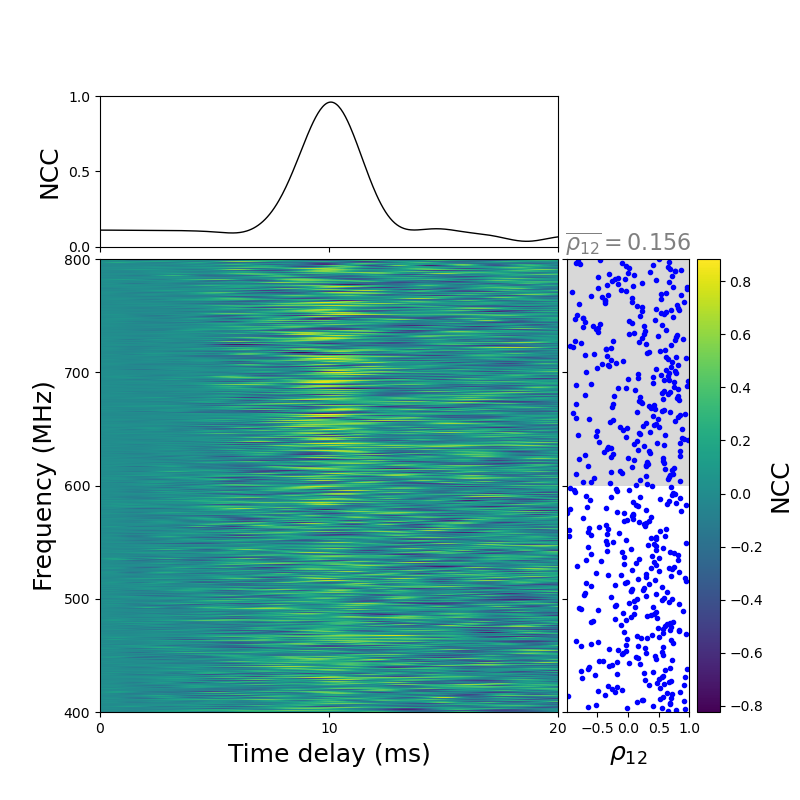}
     \caption{Same as Figure~\ref{fig3} but for a simulated unlensed FRB signal.}\label{fig4}
\end{figure*}

Furthermore, we simulate a negative case (i.e. a typical unlensed FRB signal) for comparison. In this case, the background is set as the same as the lensed signal. However, the profiles of two pulses are significantly different. The first pulse is mocked as a Gaussian profile but the second one is configured as a complicated hybrid of several Gaussian profiles. Moreover, flux ratios at different frequency bins are also set to vary randomly. The frequency-time (``waterfall") plot of the mocked unlensed signal is shown in the left panel of Figure~\ref{fig4}. Again, we first carry out NCC analysis for the mocked pulses and the result is presented in the right panel. It is found that the time delay maximizing the NCC value is significantly different for different frequency bin. It implies that a specific lensing time delay could not be derived from these two peaks. Then, PCCs for each frequency bin of this mocked signal are computed and plotted in the right panel of Figure \ref{fig4}. We obtain $\overline{\rho_{12}}=0.156$, which verifies a very weak positive correlation between these two peaks. That is, these results show strong evidence of an unlensed signal, which is in excellent agreement with the input of the simulation. Therefore, we conclude that these algorithms are robust for identifying lensing signatures with observable time delay $\gtrsim$ few milliseconds from FRB observations.

\begin{deluxetable*}{ccccc}
\tablenum{1}
\tablecaption{33 FRBs and evidence of not being lensed.}\label{tab1}
\tablewidth{0pt}
\tablehead{
\colhead{TNS name} & \colhead{Repeater source name} & \colhead{Frequency drift} & \colhead{Impossible flux ratio} & \colhead{Different structures of peaks} 
}
\startdata
FRB20180814B & {} & {} & {} & $\checkmark$\\
FRB20180917A & FRB20180814A & $\checkmark$ & {} & $\checkmark$\\
FRB20181019A & FRB20180916B & $\checkmark$ & {} & {}\\
FRB20181028A & FRB20180814A & $\checkmark$ & {} & {}\\
FRB20181104C & {} & {} & {} & $\checkmark$\\
FRB20181119D & FRB20121102A & $\checkmark$ & {} & {}\\
FRB20181125A & {} & {} & $\checkmark$ & $\checkmark$\\
FRB20181128A & FRB20181128A & $\checkmark$ & {} & $\checkmark$\\
FRB20181128C & {} & {} & $\checkmark$ & $\checkmark$\\
FRB20181222A & FRB20180916B & $\checkmark$ & {} & $\checkmark$\\
FRB20181223A & FRB20180916B & {} & {} & $\checkmark$\\
FRB20181226A & FRB20180916B & $\checkmark$ & {} & {}\\
FRB20181226B & {} & {} & $\checkmark$ & $\checkmark$\\
FRB20181228D & {} & {} & $\checkmark$ & $\checkmark$\\
FRB20190104A & {} & $\checkmark$ & {} & $\checkmark$\\
FRB20190109A & {} & {} & {} & $\checkmark$\\
FRB20190111A & {} & $\checkmark$ & {} & {}\\
FRB20190122C & {} & $\checkmark$ & {} & $\checkmark$\\
FRB20190124C & {} & {} & {} & $\checkmark$\\
FRB20190208A & FRB20190208A & $\checkmark$ & $\checkmark$ & {}\\
FRB20190213B & FRB20190212A & $\checkmark$ & {} & {}\\
FRB20190301A & FRB20190222A & $\checkmark$ & {} & $\checkmark$\\
FRB20190308C & {} & {} & $\checkmark$ & {}\\
FRB20190422A & {} & $\checkmark$ & $\checkmark$ & $\checkmark$\\
FRB20190423A & {} & $\checkmark$ & {} & {}\\
FRB20190423B & {} & {} & {} & $\checkmark$\\
FRB20190519A & FRB20180916B & $\checkmark$ & {} & $\checkmark$\\
FRB20190519B & FRB20180916B & $\checkmark$ & $\checkmark$ & $\checkmark$\\
FRB20190527A & {} & $\checkmark$ & $\checkmark$ & {}\\
FRB20190604F & FRB20180916B & $\checkmark$ & {} & $\checkmark$\\
FRB20190609A & {} & {} & {} & $\checkmark$\\
FRB20190611A & FRB20180814A & $\checkmark$ & {} & $\checkmark$\\
FRB20190625E & FRB20180814A & $\checkmark$ & $\checkmark$ & {}
\enddata
\end{deluxetable*}

\begin{deluxetable*}{ccccccc}
\tablenum{2}
\tablecaption{Normalised cross-correlation and Pearson correlation coefficient result of 12 FRBs. "Strong NCC" means that there is a strong cross-correlation at the same time delay for different frequency bins. $\overline{\rho_{12}}$ represents the average value of PCC in the frequency bins with high $\rm SNR$. }\label{tab2}
\tablewidth{0pt}
\tablehead{
\colhead{TNS name} & \colhead{Repeater of FRB} & \colhead{Strong NCC} & \colhead{$\overline{\rho_{12}}$}}
\startdata
FRB20181117B & {} & no & 0.076\\
\hline
FRB20181125A & {} & no & -0.013\\
\hline
FRB20181222E & {} & no & 0.007\\
\hline
FRB20181224E & {} & no & 0.056\\
\hline
FRB20190131D & {} & no & 0.030\\
\hline
FRB20190308B & {} & no & 0.044\\
\hline
FRB20190308C & {} & no & 0.072\\
\hline
FRB20190411C & {} & no & 0.210\\
\hline
FRB20190421A & FRB20190303A & no & 0.008\\
\hline
FRB20190501B & {} & no & -0.065\\
\hline
FRB20190601C & {} & no & -0.075\\
\hline
FRB20190605B & FRB20180916B & no & 0.102\\
\enddata
\end{deluxetable*}

\section{Results}
In this section, we first present the search result of lensing signatures from the latest FRB observations. Then, we show constraints on the abundance of PBH derived from this search result. 

\subsection{Search result of lensing signatures}
As suggested in~\citet{Munoz2016} and~\citet{Liao2020}, a lensed FRB should appear in dynamic spectrum as two pulses with almost the same profile and only different from each other by flux magnification and time delay. We have carefully checked the dynamic spectra of the latest $593$ independent FRBs. Since we target lensing signatures with observable time delay greater than the duration of the burst, we only focus on FRBs with multiple peaks. In addition to FRBs: FRB20121002A, FRB20121102A (repeating), FRB20170827A, FRB20180814A (repeating), FRB20181112A, and FRB20181123B reported in \citet{Liao2020}, we find other dozens of FRBs with double/multiple-peak structures and list them in Table~\ref{tab1} and Table~\ref{tab2}. For apparently one-off events, the most probable case is that they are intrinsic single unlensed bursts. However, in theory, there are also little chance that the first relative weak (low SNR) burst is lensed by a deflector with flux ratio greater than a threshold. In this case, the second (lensed) image might be too weak to detectable. Therefore, SNR dependent flux ratio thresholds should be used when we calculate the cross section of each FRB~\citep{Krochek2021}. Here, we take $R_{f,\rm{max}}=\rm{SNR}/10$ in our following analysis. Values of SNR for all currently available FRBs are presented in Figure~\ref{fig}.

For FRBs listed in Table~\ref{tab1}, they have been excluded as lensed candidates because of obvious characteristics which significantly violate predictions of gravitational lensing theories in their dynamic spectra, such as ``frequency drift" phenomenon, impossible flux ratio (i.e. the second pulse is brighter than the first one)~\footnote{Lensing theories predict that lens potential more complicate than the perfect spherically symmetric case may generate $>2$ images. In this case, the flux ratio criterion alone is not enough judge whether an FRB is lensed or not. Therefore, for FRB20190308C and the first two peaks of FRB20181125A, we further carefully check them with the NCC and PCC methods and results are also shown in Table~\ref{tab2}. It also suggests that there is no obvious cross-correlation caused by the lensing effect.}, and different structures of peaks. For example, the dynamic spectra of FRB20181019A with three peaks clearly show the ``frequency drift" phenomenon as mentioned in~\citet{Liao2020}. FRB20190124C is not lensed since the pulses show significant different structures. Therefore, after carefully checking characteristic features in their dynamic spectra, we find that FRBs with multiple peaks presented in Table \ref{tab1} are not lensed candidates. 

\begin{figure*}
    \centering
     \includegraphics[width=0.3\textwidth, height=0.4\textwidth]{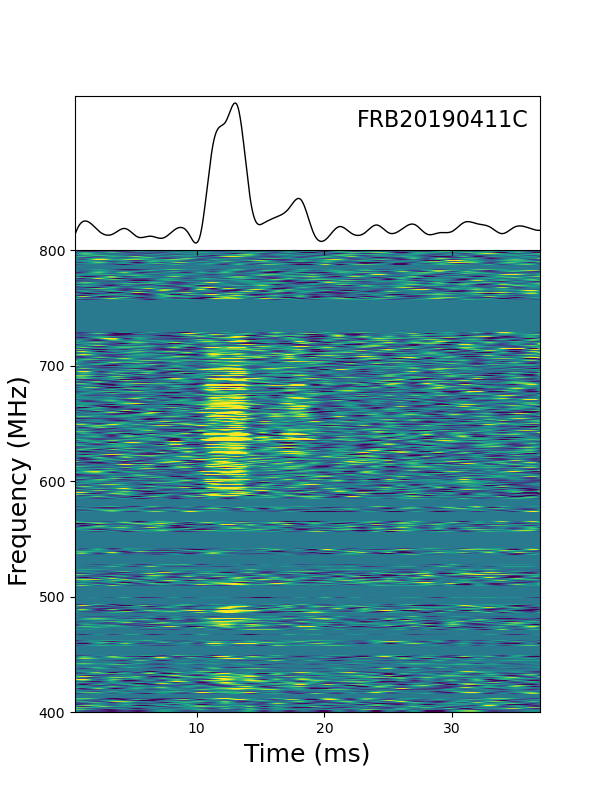}
     \includegraphics[width=0.4\textwidth, height=0.4\textwidth]{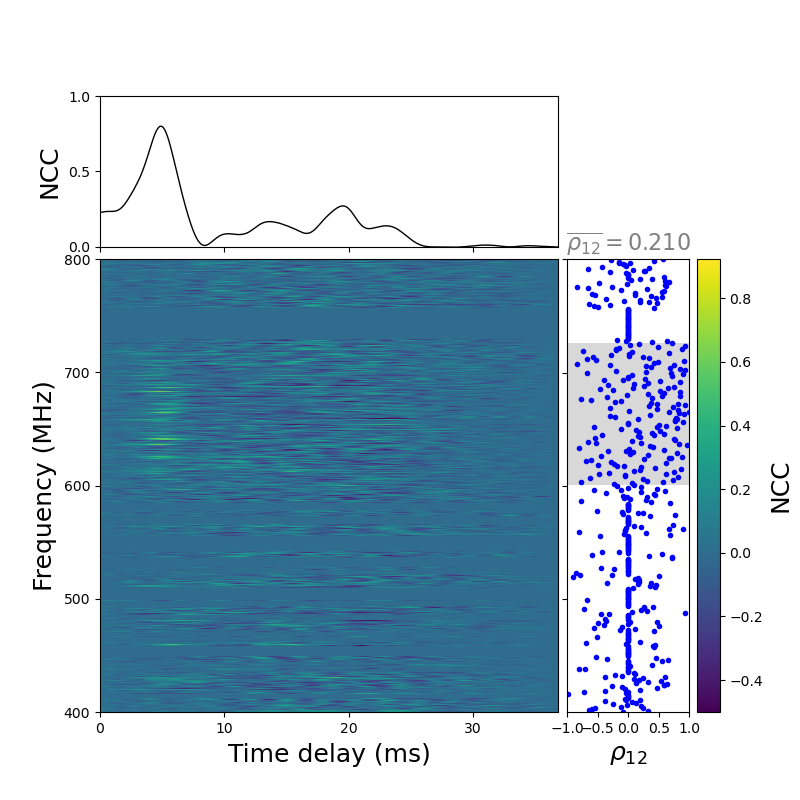}
     \caption{Same as Figure~\ref{fig3} but for FRB20190411C from the first CHIME FRB catalog.}\label{fig5}
\end{figure*}

\begin{figure*}
    \centering
     \includegraphics[width=0.3\textwidth, height=0.4\textwidth]{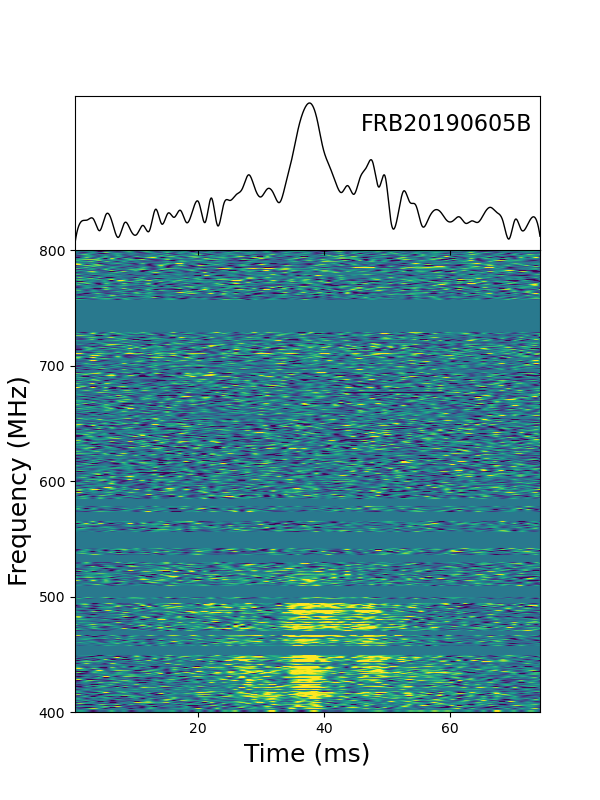}
     \includegraphics[width=0.4\textwidth, height=0.4\textwidth]{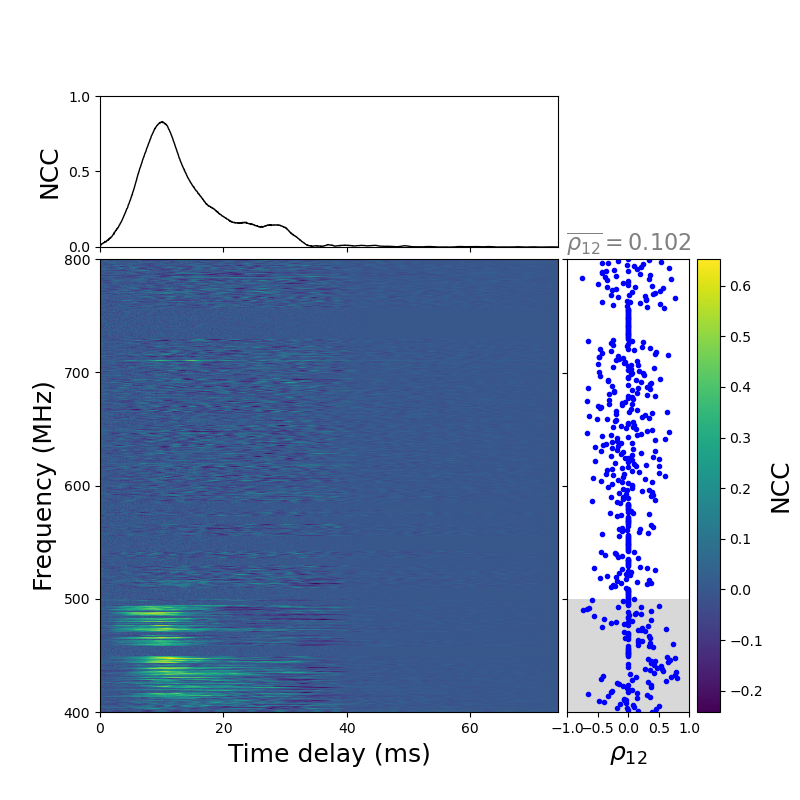}
     \caption{Same as Figure~\ref{fig3} but for FRB20190605B from the first CHIME FRB catalog.}\label{fig6}
\end{figure*}

\begin{figure}[ht!]
    \centering
     \includegraphics[width=0.4\textwidth, height=0.3\textwidth]{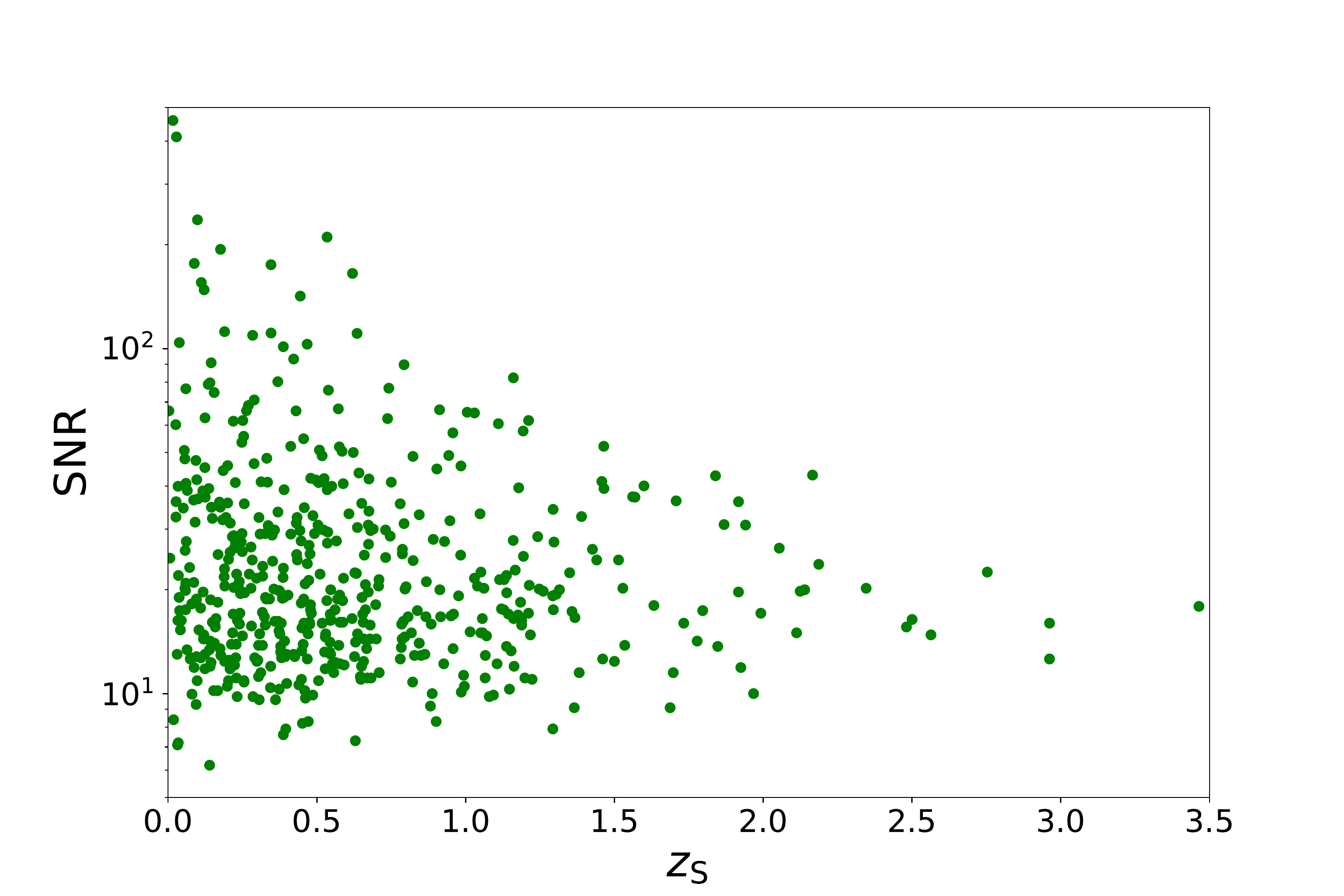}
     \caption{Two-dimensional distribution of SNR and inferred redshifts.}\label{fig}
\end{figure}

For FRBs listed in Table~\ref{tab2}, their characteristic features in dynamic spectra are not enough to identify them as lensed or unlensed candidates since pulses in each FRB show similar profile and frequency range. In addition, the second pulse of these bursts is fainter than the first corresponding pulse. This is also consistent with the gravitational lensing prediction. Therefore, we apply the NCC and PCC algorithms to identify whether these sources are lensed or unlensed. For example, the dynamic spectrum, NCC and PCC analysis results of FRB20190411C have been shown in Figure~\ref{fig5}. As suggested in the right panel, there is no strong cross-correlation between two peaks in different frequency bins and the PCC value $\overline{\rho_{12}}=0.210$ also demonstrates weak correlation between two pulses. In Figure~\ref{fig6}, we also present the dynamic spectrum, NCC and PCC analysis results of another example, FRB20190605B. There is obvious correlation between two pulses in the frequency range of 400-500 MHz. However, the maximum of NCC happens at different time delays for different frequency bins. Moreover, the PCC value $\overline{\rho_{12}}=0.102$ again verifies weak correlation between two pulses of this burst. Therefore, according to the NCC and PCC analysis results shown in Table \ref{tab2}, no strong evidence of lensed candidate has been identified in these FRBs.

\begin{figure}[ht!]
    \centering
     \includegraphics[width=0.4\textwidth, height=0.3\textwidth]{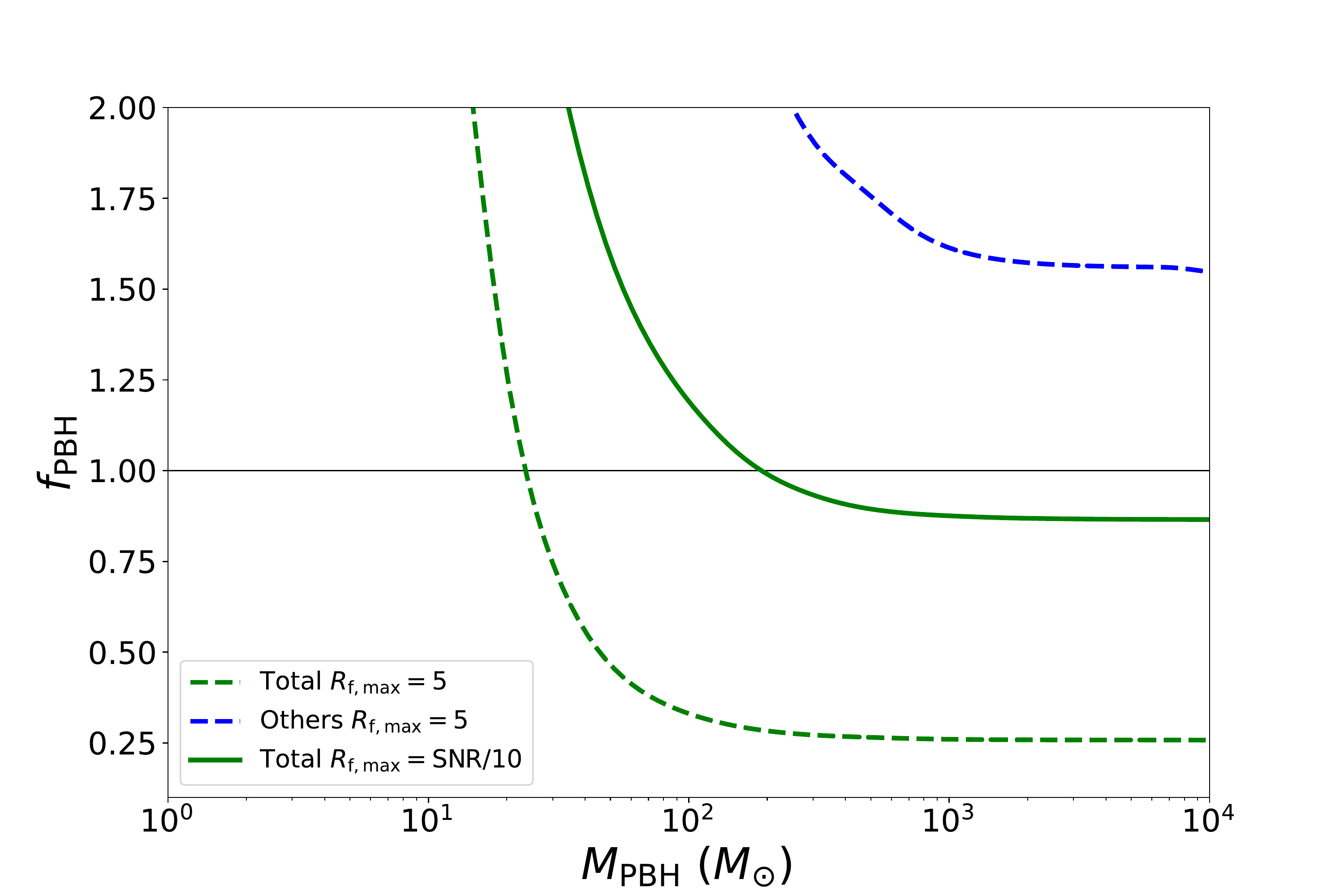}
     \caption{Constraints on the upper limits of the fraction of PBH in dark matter with a monochromatic mass function, which is based on the null search result of lensed signal in the latest FRB observation. Total and Others represent FRBs of full sample and FRBs from other facilities, respectively.}\label{fig7}
\end{figure}

\begin{figure}[ht!]
    \centering
     \includegraphics[width=0.4\textwidth, height=0.3\textwidth]{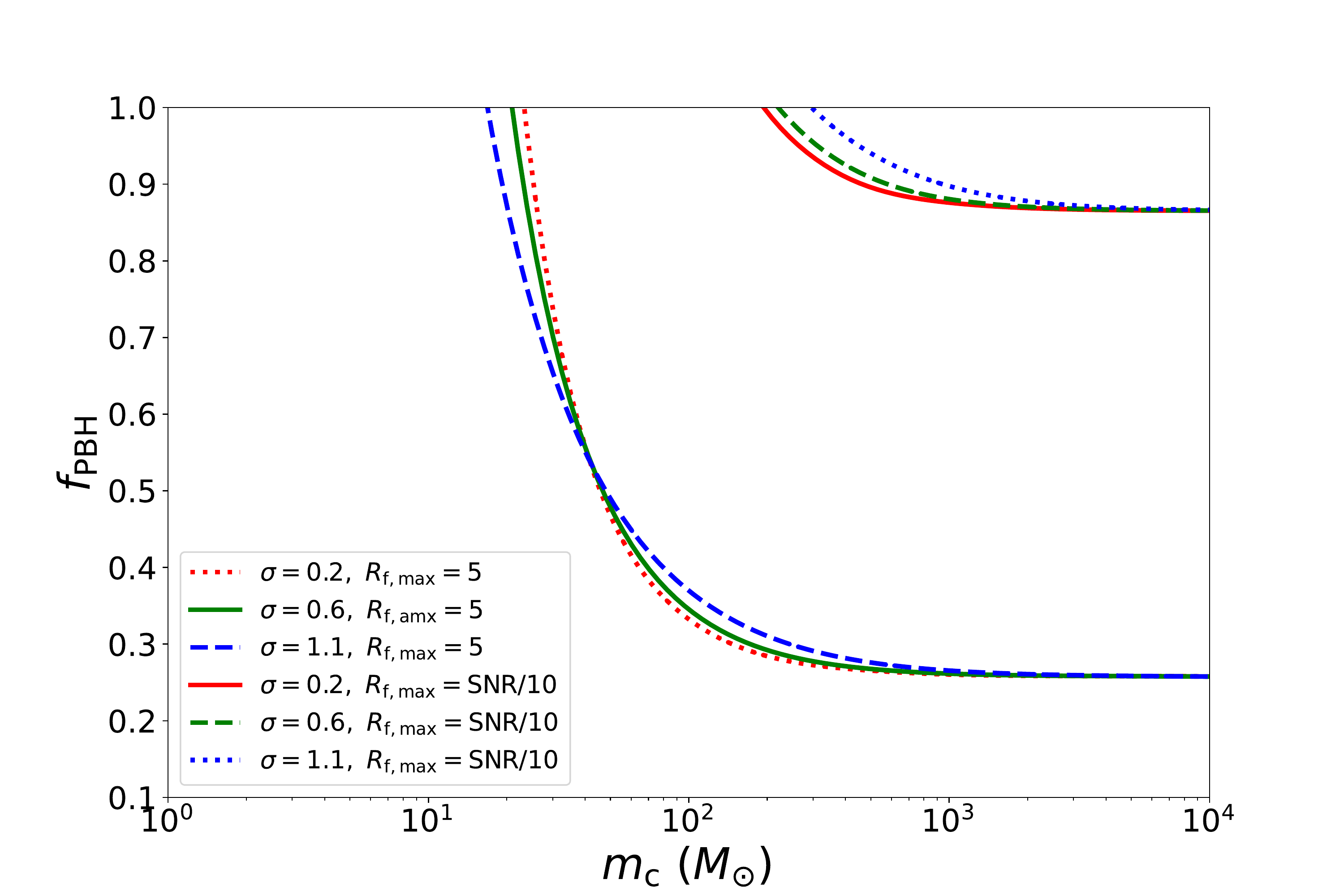}
     \caption{Same as Figure~\ref{fig7}. Constraints on the upper limits of the fraction of PBH in dark matter with a log-normal mass function, which is based on the fact that no lensed signal has been found in all currently available $593$ FRBs.}\label{fig8}
\end{figure}

\subsection{Constraints on $f_{\rm PBH}$}
As shown in Figure~\ref{fig2}, there is the two-dimensional distribution of widths and inferred redshifts. We can find that the inferred redshift of FRBs tend to be concentrated at low redshifts $z<1.5$ in Figure~\ref{fig1}, which will be one of the factors that have an important influence on the results. We follow the standard operating procedure in this paper for constraints on the the abundance of PBHs. For the monochromatic mass function, each $(M_{\rm PBH},f_{\rm PBH})$ corresponds to an expected number of lensed FRB signals according to Equation~\ref{eq6} and Equation~\ref{eq9}. Since no lensed signal has been found in the current data, the curve in the $(M_{\rm PBH},f_{\rm PBH})$ parameter space that predicts at least three detectable lensed signal should be ruled out at $95\%$ confidence level. As shown in Figure~\ref{fig7}, the mass can be tested down to $\sim30~M_{\odot}$ and $f_{\rm PBH}$ is gradually constrained to $87\%$ for the mass range $\gtrsim500~M_{\odot}$ at $95\%$ confidence level by using SNR dependent flux ratio thresholds. It should be presented that these constraints could become three times stronger when a commonly used constant flux ratio threshold $R_{f,\rm{max}}=5$ is considered. In Figure~\ref{fig7}, we also show the result of constraints on $f_{\rm PBH}$ from the FRB observations without CHIME. It suggests that the first CHIME FRB catalog has greatly improved the constraints on the abundance of PBH. Although current constrains are relative weak, especially for small masses, there will be much better constraints from large number of FRBs with high SNR in the near future~\citep{Liao2020, Laha2020, Munoz2016}. 

 Recently, constraints on PBHs in the stellar mass range with the Bayesian inference method and the log-normal distribution from the GWTC-1/GWTC-2 GW catalog have been extensively studied~\citep{Chen2019a,Wu2020,Gert2020,Kaze2020,Luca2020}. Therefore, in order to compare constraints on PBH from FRB observations with those from GW detections in this intriguing mass window, we next derive projected constraints on $f_{\rm PBH}$ from the latest FRB observations assuming that the mass of PBHs follows a log-normal distribution. Results are shown in Figure~\ref{fig8}. we assume the parameter $\sigma=0.2$, $0.6$, and $1.1$ in log-normal mass function. Analogously, each $(m_{\rm c},f_{\rm PBH})$ corresponds to an expected number of lensed FRB signals according to Equation~\ref{eq9} and Equation~\ref{eq11}. Since no lensed signal has been found in the current data, the region above the curve that predicts at least one detectable lensed signal in the $(m_{\rm c},f_{\rm PBH})$ parameter space should be ruled out. We find our results at $m_{\rm c}\leq100~M_{\odot}$ still weaker compared with the results of GW~\citep{Luca2020, Gert2020}.

\section{Conclusions and discussions}
FRBs are one of the most mysterious phenomena in astrophysics. Although we do not yet understand formation mechanisms of these bursts, some their unique observational features make them as promising probes for astrophysical and cosmological purposes. Meanwhile, the recent detection of GWs from mergers of binary stellar mass black holes has stimulated great interests in PBHs of this mass range. Gravitational lensing effect of transients with millisecond duration and high event rate, e.g. FRBs, has been put forward as one of the cleanest probes for exploring properties of PBHs in the mass range $1-100~M_{\odot}$. In this paper, we first propose to use the normalised cross-correlation and Pearson correlation coefficient algorithms for identifying lensed signatures in FRB observations. Next, we have carefully checked all events with multiple peaks by comparing their observational properties, such as dynamic spectra, morphologies of the pulses, and the flux ratio, with lensing theory predictions. Most of them have been ruled out as lensed FRB candidate. For the rest $\sim10$ FRBs whose waterfalls apparently look like lensed events, we further apply the NCC and PCC algorithms to testify the evidence of them as lensed signals. The tests suggest that there is no strong evidence of correlation between peaks of these FRBs. As a result, we conclude that there is no lensing candidate with time delay greater than duration in currently available FRB observations.

On the basis of the null search result, we derive direct constraints on the abundance of PBH from the latest FRB observations and obtain that, for a monochromatic mass distribution, the abundance of PBHs is constrained to $\leq87\%$ and $\leq26\%$ when an SNR dependent and a constant flux ratio threshold is considered, respectively. This constraints has been significantly improved owing to the inclusion of the first CHIME FRB catalog. Moreover, we also investigate constraints by considering a log-normal mass distribution, which is naturally predicted by some popular inflation models, to compare results constrained from FRB observations with those from GW detections. Although this constraint is weaker compared with the results from GW, it will be significantly improved with the rapid increase of the number of FRBs detected by wide-field surveys (like CHIME and DAS-2000) in the near future. As a result, there would be significant overlap between areas constrained from FRB observations and the ones from GW detection in the parameter space of the mass distribution. Then, it would be  possible to jointly constrain the abundance and mass distribution of PBHs by combining this two kinds of  promising multi-messenger observations. It is foreseen that these joint constraints will of great importance for exploring the nature of PBHs in the stellar mass range, or even their formation mechanisms relating to physics of the early universe.

\section{Acknowledgements}
The authors thank the Yukawa Institute for Theoretical Physics at Kyoto University, where this work was initiated  during the YITP International Molecule-type Workshop ``Fast Radio Bursts: A Mystery Being Solved?". We would like to thank Ziggy Pleunis for helpful discussions. We are also very grateful to the referee for the valuable comments, which have allowed us to improve our manuscript. This work was supported by the National Natural Science Foundation of China under Grants Nos. 11920101003, 11722324, 11603003, 11633001, 11973034 and U1831122, the Strategic Priority Research Program of the Chinese Academy of Sciences, Grant No. XDB23040100, and the Interdiscipline Research Funds of Beijing Normal University.

\end{document}